\documentclass[12pt,prd,noshowpacs,nofootinbib,notitlepage]{revtex4-1}
\usepackage{setspace}
\usepackage{graphicx,color}
\usepackage{maplestd2e}
\usepackage{bm}

\DeclareSymbolFontAlphabet{\mathcal}{symbols}
\DeclareSymbolFont{symbols}{OMS}{xmdcmsy}{m}{n}
\DeclareSymbolFont{largesymbols}{OMX}{xmdcmex}{m}{n}
\SetSymbolFont{symbols}{bold}{OMS}{xmdcmsy}{b}{n}
\begin{document}  
\title{\color{blue}\Large Nuclear-size effects\\ and a numerical approach to the Dirac equation}
\author{Bob Holdom}
\email{bob.holdom@utoronto.ca}
\affiliation{Department of Physics, University of Toronto, Toronto, Ontario, Canada  M5S1A7}
\author{Roman Koniuk}
\email{koniuk@yorku.ca}
\affiliation{Department of Physics, York University, Toronto, Ontario, Canada  M3J1P3}
\begin{abstract}
Due to some current interest in this subject we have produced this note. There is no claim to anything new, except possibly to show that a direct numerical approach is quite simple and instructive. For comparison purposes we include a section on the Coulomb Klein-Gordon equation.
\end{abstract}
\maketitle

\section{Review of Dirac equation and perturbation theory}

The problem of a single electron in a central field is thoroughly discussed in Bjorken and Drell \cite{BandD}. We also adopt some notation from \cite{Friar:1978wv}. The Hamiltonian
\begin{equation}
H\psi =[\bm{\alpha\cdot p}+ \beta m +V(r)]\psi = E\psi
\end{equation}
commutes with the total angular momentum
\begin{equation}
\bm{J} =\bm{ L} +\bm{S} = \bm{r\times p} + \frac{\bm\sigma}{2}.
\end{equation}
The  four-component spinor $\psi$ is constructed to be a simultaneous eigenfunction of $H, J^2$ and $J_z$. It is convenient to write the general solution for a given $j,m$ as
\begin{equation}
\psi_{jm} = 
\left(
\begin{array}{c}\displaystyle
\frac{f(r)}{r}\chi_{jm}^{(\pm)}   \\\displaystyle
-i\frac{g(r)}{r} \bm{\sigma\cdot{ \hat{r}}}\,\chi_{jm}^{(\pm)}
\end{array}
\right).
\end{equation}
The parity $(\pm)$ refers to solutions with $j=l\pm\frac{1}{2}$ where $j\geq \frac{1}{2}$.
The two-component spinors $\chi^{(\pm)}$ are eigenstates of an auxiliary operator $K= -(1+\bm{\sigma\cdot{L}})$ such that $K\chi = \kappa\chi$ with $\kappa = \mp(j+\frac{1}{2})$ for $j=l\pm\frac{1}{2}$.
With this notation the Dirac equation can be reduced to the following radial equations,
\begin{eqnarray}
\frac{df(r)}{dr} + \frac{\kappa}{r}f(r) - (E +m -V(r))g(r) &=& 0,\nonumber\\
\frac{dg(r)}{dr} - \frac{\kappa}{r}g(r) + (E -m -V(r))f(r) &=& 0.
\label{e3}\end{eqnarray}

For the point charge potential $V(r) = -\alpha/r$ the solutions are known. The energy eigenvalues are
\begin{equation}
E_n= m\left[1+\left(\frac{\alpha}{n-(j+\frac{1}{2})+\sqrt{(j+\frac{1}{2})^2-\alpha^2}}\right)^2\right]^{-1/2}
\end{equation}
where $n$ is a positive integer and the angular momentum eigenvalues $j$ range from $\frac{1}{2}$ to $j+\frac{1}{2}\le n$. The ground-state has $l=0$ which implies $j=\frac{1}{2}$ and $\kappa=-1$, and $n=1$ which implies the solution has zero nodes. Its energy is $E=m\gamma$ where $\gamma = \sqrt{1-\alpha^2}$ and the corresponding solution is
\begin{eqnarray}
\frac{f(r)}{r} &=& (2m\alpha)^{3/2}\sqrt{\frac{1+\gamma}{{2\Gamma(1+2\gamma)}}}(2m\alpha r)^{\gamma-1}e^{-m\alpha r},\nonumber\\
\frac{g(r)}{r} &=& \frac{(1-\gamma)}{\alpha}\frac{f(r)}{r}.
\label{ee1}\end{eqnarray}
For this case $\chi_{\frac{1}{2}m}^+\to\chi^{m}/\sqrt{4\pi}$ where the $\chi^m$ is the usual up or down two-component spinor for $m=\pm\frac{1}{2}$. Also note that there is another solution to the equations where $f(r)/r \sim r^{-\gamma-1}$ as $r\to0$ rather than the $r^{\gamma-1}$ behavior in (\ref{ee1}). But that solution is not normalizable.

We next consider the effect that the finite nuclear size has on the energy of the $l=0$ states for any $n\geq1$. Friar \cite{Friar:1978wv} obtained this in perturbation theory for a general nuclear charge distribution, and we summarize these results in the Appendix. He also considered various examples; here we focus on the uniformly charged sphere. Friar expresses the shift in energy due to the finite size as 
\begin{equation}
\Delta E= -\frac{(Z\alpha)^2\mu}{2}\delta_B,
\end{equation}
where $\mu$ is the reduced mass $Z$ is the nuclear charge. He then obtains
\begin{equation}
\delta_B = \frac{\xi^2}{n^3}\sum_{i=0}^2\delta_i\xi^i + \frac{\delta\xi^2}{n^3}\Delta_0^R
\end{equation}
where $\xi = Z\alpha\mu R$, $\delta =(Z\alpha)^2$ and the $\delta_i$ are
\begin{eqnarray}
\delta_0 &=& -\frac{4}{5}\\
\delta_1 &=& \frac{64}{63}\\
\delta_2 &=& -\frac{56954}{225225} + \frac{8}{25n}-\frac{2}{35n^2}-\frac{8}{25}\left(\psi(n) + 2\gamma + \log\left(\frac{2\xi}{n}\right)\right)\\
\Delta_0^R &=& \frac{4}{5}(\psi(n)  + \log(2\xi/n)+ 2\gamma)-\frac{4}{5n} + \frac{9}{5n^2} - \frac{45394}{17325}.
\end{eqnarray}
We shall stay in the infinite nuclear mass limit where the reduced mass $\mu$ can be replaced by $m$. To get some sense of the relative size of the various perturbative contributions to $\Delta E$ we give some numerical values in Table~\ref{t1}. We consider the electron and muon masses for $m$ and two choices of the proton charge radius $r_p$.

\begin{table}[htp]
\begin{center}
\begin{tabular}{||c|c|c|c|c||}
\hline
perturbative order & electron (0.84) &   muon (0.84) & \ \ \ electron  (0.88) \ \ \  & muon (0.88) \\
\hline
\hline
$1^{st}$  &$-4.57105\times10^{-6}$ & $-40.408$& $-5.01675\times10^{-6}$ &  $-44.348$\\ 
\hline
$2^{nd}$ &  $1.18951\times10^{-10}$ & $0.217422$ &  $1.36766\times10^{-10}$ &  $0.249985$\\ 
\hline
$3^{rd}$ & $7.33785\times10^{-15}$&  $0.00122601$& $8.79555\times10^{-15}$ &  $0.00146049$\\ 
\hline
relativistic &$-2.8115\times10^{-9}$ & $-0.0133812$ & $-3.07321\times10^{-9}$ &  $-0.0145761$\\ 
\hline
& & & &\\
\hline 
\hline
total correction  & $-4.57374\times10^{-6}$ & $-40.2027$  & $-5.01969\times10^{-6}$ &  $-44.1112$\\ 
\hline
\end{tabular}
\end{center}
\caption{Contributions to $\Delta E$ in meV with $Z=n=1$ and for $r_p=0.84$ and $0.88$.}
\label{t1}
\end{table}%

We may also comment on the $r=0$ boundary condition for the finite size charge where the potential is no longer singular. Now the two apparent $\ell=0$ solutions behave like $f(r)/r\sim \textrm{constant}$ or $1/r$ respectively as $r\to0$. Both are normalizable but the second one has another problem.  As can be seen in the following section, $f(r)/r$ satisfies an equation with terms that correspond to the radial laplacian. But a laplacian acting on a $1/r$ wave function produces a $\delta$-function. This means that this apparent second solution is in fact not a solution.

\section{Numerical approach using Maple}
\begin{maplegroup}
\begin{Maple Normal}{We need an environment where a differential equation can be solved to high precision and where this equation can involve a piecewise defined function. Maple is such an environment, and here we will make our approach explicit by giving the Maple code.}\end{Maple Normal}

\end{maplegroup}
\begin{maplegroup}
\begin{mapleinput}
\mapleinline{active}{1d}{Digits := 20:
}{}
\end{mapleinput}
\begin{mapleinput}
\mapleinline{active}{1d}{st1 := method = ck45, abserr = 10\symbol{94}(-15), relerr = 10\symbol{94}(-15), maxfun = 100000:
}{}
\end{mapleinput}
\end{maplegroup}
\begin{maplegroup}
\begin{Maple Normal}{The radius \mapleinline{inert}{2d}{a}{$\displaystyle a$}
of a uniformly charged sphere in 1/MeV based on \mapleinline{inert}{2d}{r__p = .88}{$\displaystyle r_{p}\approx 0.88$}
fm is given.}\end{Maple Normal}

\end{maplegroup}
\begin{maplegroup}
\begin{mapleinput}
\mapleinline{active}{1d}{l1 := \{a = .88*sqrt(5./3.)/197.3, alpha = 1/137.035999\}:
}{}
\end{mapleinput}
\begin{mapleinput}
\mapleinline{active}{1d}{mmu := 105.65837: me := .51099894:
}{}
\end{mapleinput}
\end{maplegroup}
\begin{maplegroup}
\begin{Maple Normal}{We want to compare to the Friar result for the energy shift due to the finite size effect. This is his result for the \mapleinline{inert}{2d}{n = 1}{$\displaystyle n=1$} ground state with $Z=1$
.}\end{Maple Normal}

\end{maplegroup}
\begin{maplegroup}
\begin{mapleinput}
\mapleinline{active}{1d}{xi := alpha*m*a:
DD := alpha\symbol{94}2: 
D0 := -4/5: 
D1 := 64/63:
D2 := 2248/225225-8/25*(ln(2*xi)+gamma):
CD := -28069/17325+4/5*(ln(2*xi)+gamma):
DB := D0*xi\symbol{94}2+D1*xi\symbol{94}3+D2*xi\symbol{94}4+CD*DD*xi\symbol{94}2:
EB := -(1/2)*alpha\symbol{94}2*m*DB:
}{}
\end{mapleinput}
\end{maplegroup}
\begin{maplegroup}
\begin{Maple Normal}{Here are these shifts for the muon and the electron in MeV.}\end{Maple Normal}

\end{maplegroup}
\begin{maplegroup}
\begin{mapleinput}
\mapleinline{active}{1d}{subs(m = mmu, l1, EB):q1:=evalf(
}{}
\end{mapleinput}
\mapleresult
\begin{maplelatex}
 \mapleinline{inert}{2d}{q1 := 4.4123175857857733137*10^(-8)}{}{\[\displaystyle {\it q1}\, := \, 0.000000044123175857857733137\]}
\end{maplelatex}
\end{maplegroup}
\begin{maplegroup}
\begin{mapleinput}
\mapleinline{active}{1d}{subs(m = me, l1, EB): q2:=evalf(
}{}
\end{mapleinput}
\mapleresult
\begin{maplelatex}
 \mapleinline{inert}{2d}{q2 := 5.0210593674052689345*10^(-15)}{}{\[\displaystyle {\it q2}\, := \,{ 5.0210593674052689345\times 10^{-15}}\]}
\end{maplelatex}
\end{maplegroup}
\begin{maplegroup}
\begin{Maple Normal}{The following ratio then gives the residual mass dependence of these shifts beyond the trivial \mapleinline{inert}{2d}{m^3}{$\displaystyle {m}^{3}$}
dependence.}\end{Maple Normal}

\end{maplegroup}
\begin{maplegroup}
\begin{mapleinput}
\mapleinline{active}{1d}{q1/q2*(me/mmu)\symbol{94}3;
}{}
\end{mapleinput}
\mapleresult
\begin{maplelatex}
 \mapleinline{inert}{2d}{.99407622263401855408}{}{\[\displaystyle  0.99407622263401855408\]}
\end{maplelatex}
\end{maplegroup}
\begin{maplegroup}
\begin{Maple Normal}{We want to obtain this same ratio by numerically solving the Dirac equation. We consider the point charge potential,}\end{Maple Normal}

\end{maplegroup}
\begin{maplegroup}
\begin{mapleinput}
\mapleinline{active}{1d}{V1:=(alpha, a, r) ->-alpha/r ;
}{}
\end{mapleinput}
\mapleresult
\begin{maplelatex}
 \mapleinline{inert}{2d}{V1 := proc (alpha, a, r) options operator, arrow; -alpha/r end proc}{}{\[\displaystyle {\it V1}\, := \,( {\alpha,a,r} )\mapsto -{\frac {\alpha}{r}}\]}
\end{maplelatex}
\end{maplegroup}
\begin{maplegroup}
\begin{Maple Normal}{and the potential for the uniformly charge sphere with radius \mapleinline{inert}{2d}{a}{$\displaystyle a$}
.}\end{Maple Normal}

\end{maplegroup}
\begin{maplegroup}
\begin{mapleinput}
\mapleinline{active}{1d}{V2 :=(alpha, a, r) -> piecewise(r < a, (1/2)*alpha*(r\symbol{94}2/a\symbol{94}2-3)/a, -alpha/r);
}{}
\end{mapleinput}
\mapleresult
\begin{maplelatex}
 \mapleinline{inert}{2d}{V2 := proc (alpha, a, r) options operator, arrow; piecewise(r < a, (1/2)*alpha*(r^2/a^2-3)/a, -alpha/r) end proc}{}{\[\displaystyle {\it V2}\, := \,( {\alpha,a,r} )\mapsto \cases{1/2\,{\frac {\alpha}{a} \left( {\frac {{r}^{2}}{{a}^{2}}}-3 \right) }&$r<a$\cr -{\frac {\alpha}{r}}&otherwise\cr}\]}
\end{maplelatex}
\end{maplegroup}
\begin{maplegroup}
\begin{mapleinput}
\mapleinline{active}{1d}{plot(V2(1, 1, r), r = 0 .. 5);
}{}
\end{mapleinput}
\mapleresult
\mapleplot{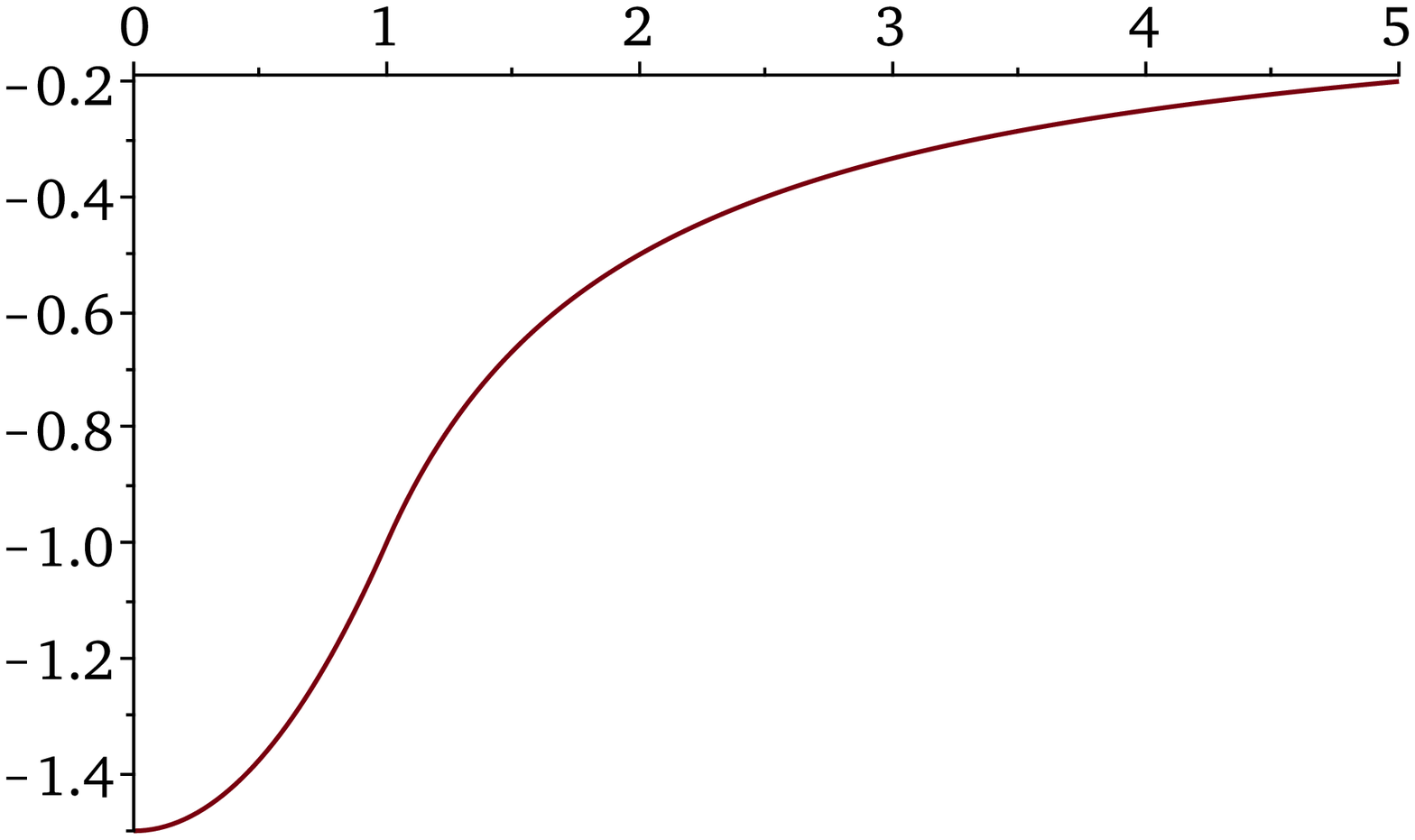}
\end{maplegroup}
\begin{maplegroup}
\begin{Maple Normal}{The Dirac equation for the ground state reduces to following equations for \mapleinline{inert}{2d}{f(r)}{$\displaystyle f \left( r \right) $}
and \mapleinline{inert}{2d}{g(r)}{$\displaystyle g \left( r \right) $}
. (Maple will interpret these expressions as equations.)}\end{Maple Normal}

\end{maplegroup}
\begin{maplegroup}
\begin{mapleinput}
\mapleinline{active}{1d}{e1:=diff(f(r), r)-f(r)/r-(E+m-V(alpha, a, r))*g(r);
}{}
\end{mapleinput}
\mapleresult
\begin{maplelatex}
 \mapleinline{inert}{2d}{e1 := diff(f(r), r)-f(r)/r-(E+m-V(alpha, a, r))*g(r)}{}{\[\displaystyle {\it e1}\, := \,{\frac {\rm d}{{\rm d}r}}f \left( r \right) -{\frac {f \left( r \right) }{r}}- \left( E+m-V \left( \alpha,a,r \right)  \right) g \left( r \right) \]}
\end{maplelatex}
\end{maplegroup}
\begin{maplegroup}
\begin{mapleinput}
\mapleinline{active}{1d}{e2:=diff(g(r), r)+g(r)/r+(E-m-V(alpha, a, r))*f(r);
}{}
\end{mapleinput}
\mapleresult
\begin{maplelatex}
 \mapleinline{inert}{2d}{e2 := diff(g(r), r)+g(r)/r+(E-m-V(alpha, a, r))*f(r)}{}{\[\displaystyle {\it e2}\, := \,{\frac {\rm d}{{\rm d}r}}g \left( r \right) +{\frac {g \left( r \right) }{r}}+ \left( E-m-V \left( \alpha,a,r \right)  \right) f \left( r \right) \]}
\end{maplelatex}
\end{maplegroup}
\begin{maplegroup}
\begin{Maple Normal}{We convert these into a second order equation.}\end{Maple Normal}

\end{maplegroup}
\begin{maplegroup}
\begin{mapleinput}
\mapleinline{active}{1d}{isolate(e1, g(r)):
subs(
numer(
e3:=simplify(
}{}
\end{mapleinput}
\mapleresult
\begin{maplelatex}
 \mapleinline{inert}{2d}{e3 := r*(E+m-V(alpha, a, r))*(diff(f(r), r, r))+(r*(diff(f(r), r))-f(r))*(diff(V(alpha, a, r), r))+r*f(r)*(E-m-V(alpha, a, r))*(E+m-V(alpha, a, r))^2}{}{\[\displaystyle {\it e3}\, := \,r \left( E+m-V \left( \alpha,a,r \right)  \right) {\frac {{\rm d}^{2}}{{\rm d}{r}^{2}}}f \left( r \right) + \left( r{\frac {\rm d}{{\rm d}r}}f \left( r \right) -f \left( r \right)  \right) {\frac {\partial }{\partial r}}V \left( \alpha,a,r \right) \\
\mbox{}+rf \left( r \right)  \left( E-m-V \left( \alpha,a,r \right)  \right)  \left( E+m-V \left( \alpha,a,r \right)  \right) ^{2}\]}
\end{maplelatex}
\end{maplegroup}
\begin{maplegroup}
\begin{Maple Normal}{Then for the point charge the equation is}\end{Maple Normal}

\end{maplegroup}
\begin{maplegroup}
\begin{mapleinput}
\mapleinline{active}{1d}{e4:=subs(V = V1, e3):
}{}
\end{mapleinput}
\end{maplegroup}
\begin{maplegroup}
\begin{Maple Normal}{We check the exact solution and the corresponding energy.}\end{Maple Normal}

\end{maplegroup}
\begin{maplegroup}
\begin{mapleinput}
\mapleinline{active}{1d}{f(r) = r\symbol{94}sqrt(-alpha\symbol{94}2+1)*exp(-m*alpha*r), E = m*sqrt(-alpha\symbol{94}2+1);
subs(
}{}
\end{mapleinput}
\mapleresult
\begin{maplelatex}
 \mapleinline{inert}{2d}{f(r) = r^sqrt(-alpha^2+1)*exp(-m*alpha*r), E = m*sqrt(-alpha^2+1)}{}{\[\displaystyle f \left( r \right) ={r}^{ \sqrt{-{\alpha}^{2}+1}}{{\rm e}^{-m\alpha\,r}},\,E=m \sqrt{-{\alpha}^{2}+1}\]}
\end{maplelatex}
\mapleresult
\begin{maplelatex}
 \mapleinline{inert}{2d}{0}{}{\[\displaystyle 0\]}
\end{maplelatex}
\end{maplegroup}
\begin{maplegroup}
\begin{Maple Normal}{For the uniformly charge sphere the equation to solve is the following.}\end{Maple Normal}

\end{maplegroup}
\begin{maplegroup}
\begin{mapleinput}
\mapleinline{active}{1d}{e5:=subs(V = V2, e3):
}{}
\end{mapleinput}
\end{maplegroup}
\begin{maplegroup}
\begin{Maple Normal}{Since Maple can handle piecewise functions there is no need to do matching across the boundary \mapleinline{inert}{2d}{r = a}{$\displaystyle r=a$}. So we numerically integrate this equation from the origin with boundary conditions \mapleinline{inert}{2d}{f(0) = 0}{$\displaystyle f \left( 0 \right) =0$}
and \mapleinline{inert}{2d}{(D(f))(0) = 1}{$\displaystyle \mbox {D} \left( f \right)  \left( 0 \right) =1$}
. We are not interested in the normalization of \mapleinline{inert}{2d}{f(r)}{$\displaystyle f \left( r \right) $}
. We adjust \mapleinline{inert}{2d}{E}{$\displaystyle E$}
via the shooting method to obtain the zero nodes solution with \mapleinline{inert}{2d}{f(infinity) = 0}{$\displaystyle f \left( \infty  \right) =0$}
. For the electron case:}\end{Maple Normal}

\end{maplegroup}
\begin{maplegroup}
\begin{mapleinput}
\mapleinline{active}{1d}{Ee:=.5109853341259963716:
ip := 0: ic := \{f(ip) = ip, (D(f))(ip) = 1\}:
eq := \{subs(m = me, l1, E = Ee, e5)\}:
s1 := dsolve(eq union ic, \{f(r)\}, type = numeric, st1):
odeplot(s1, [r, f(r)], ip .. 9000);
}{}
\end{mapleinput}
\mapleresult
\mapleplot{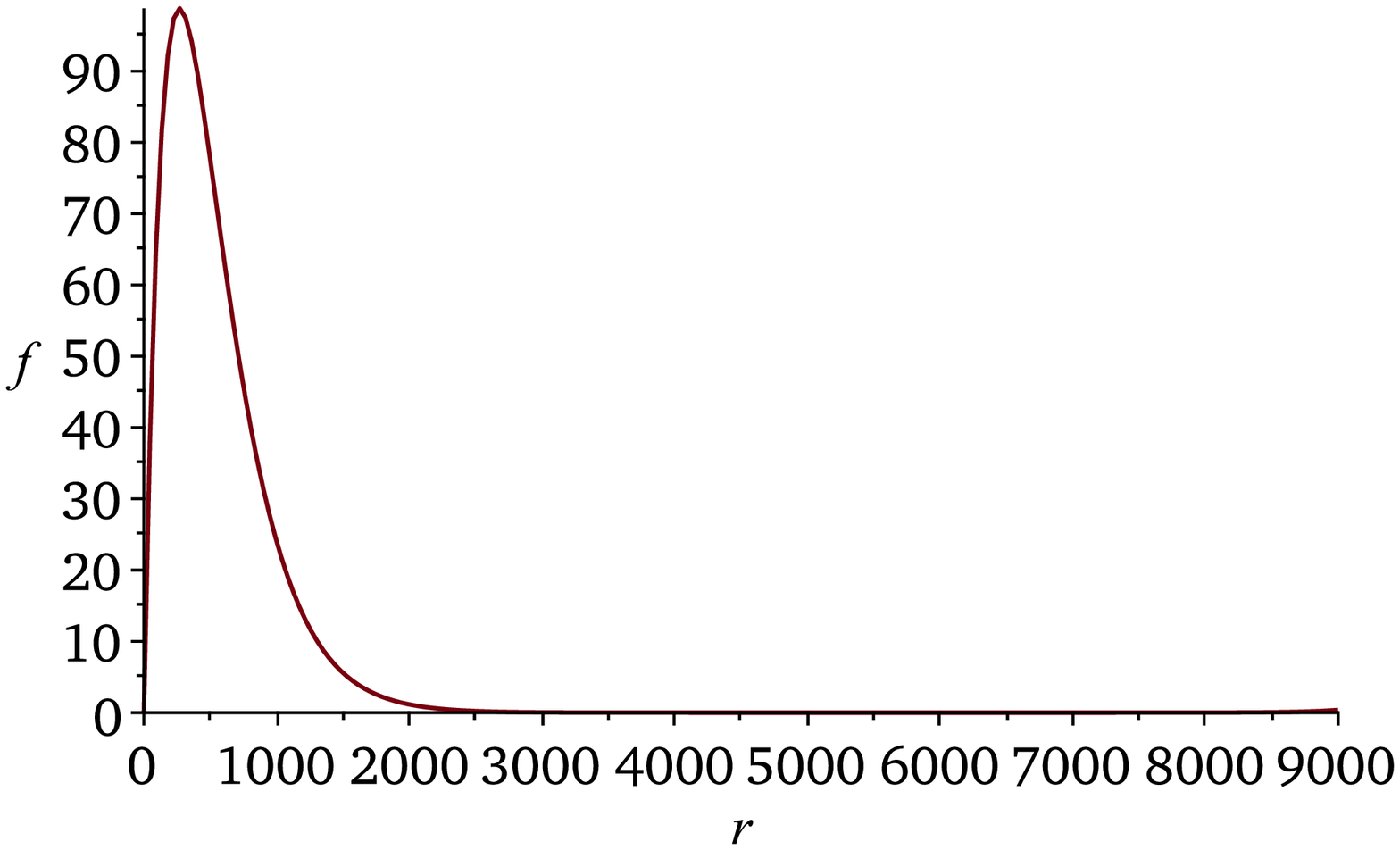}
\end{maplegroup}
\begin{maplegroup}
\begin{Maple Normal}{For the muon:}\end{Maple Normal}

\end{maplegroup}
\begin{maplegroup}
\begin{mapleinput}
\mapleinline{active}{1d}{Emu:=105.655556781007189:
ip := 0: ic := \{f(ip) = ip, (D(f))(ip) = 1\}:
eq := \{subs(m = mmu, l1, E = Emu, e5)\}:
s1 := dsolve(eq union ic, \{f(r)\}, type = numeric, st1):
odeplot(s1, [r, f(r)], ip .. 38);
}{}
\end{mapleinput}
\mapleresult
\mapleplot{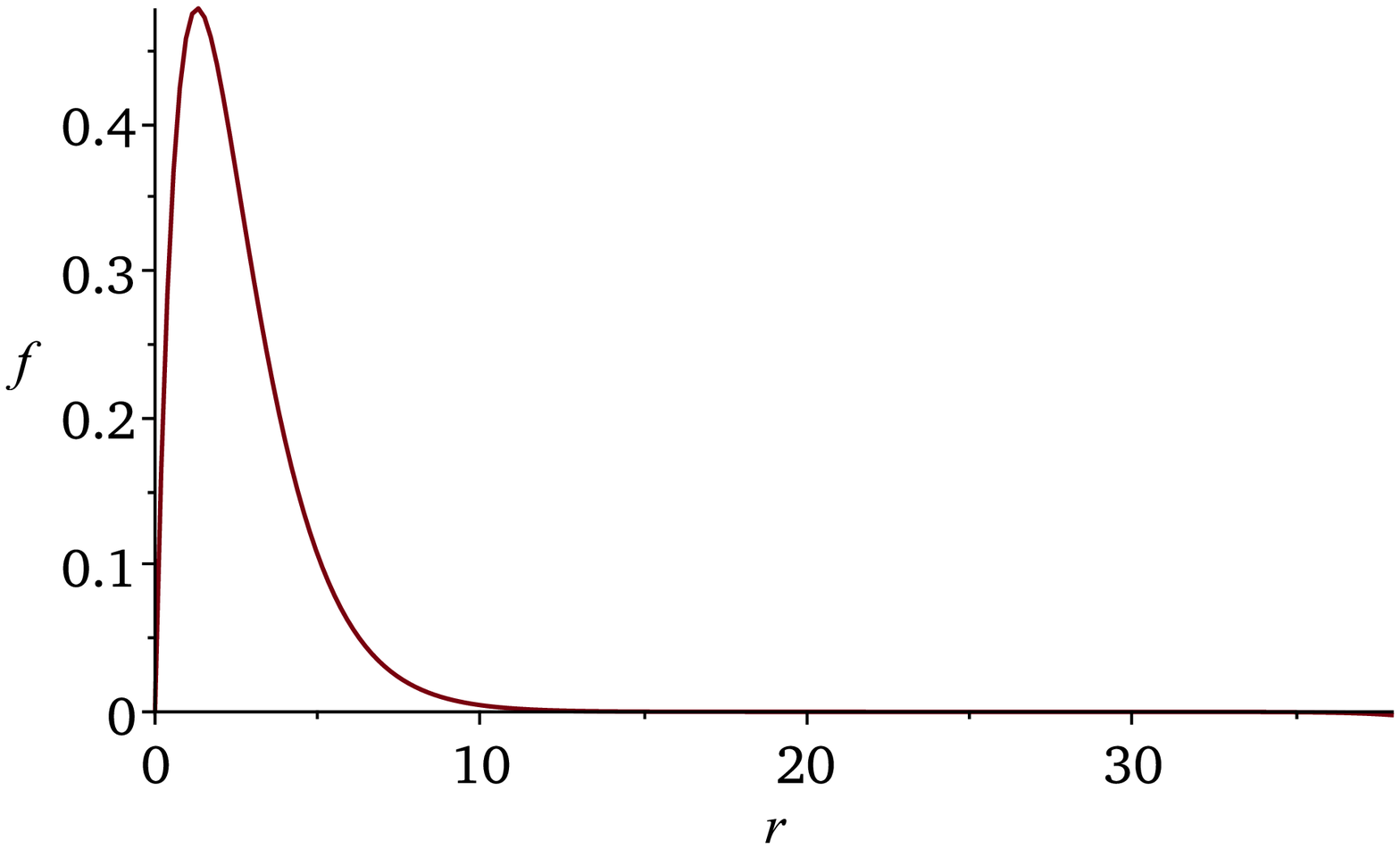}
\end{maplegroup}
\begin{maplegroup}
\begin{Maple Normal}{We need to compare these energies to the exact energies for the point charge case.}\end{Maple Normal}

\end{maplegroup}
\begin{maplegroup}
\begin{mapleinput}
\mapleinline{active}{1d}{q3:=subs(m = mmu, l1, m*sqrt(-alpha\symbol{94}2+1));
}{}
\end{mapleinput}
\mapleresult
\begin{maplelatex}
 \mapleinline{inert}{2d}{q3 := 105.65555673688407403}{}{\[\displaystyle {\it q3}\, := \, 105.65555673688407403\]}
\end{maplelatex}
\end{maplegroup}
\begin{maplegroup}
\begin{mapleinput}
\mapleinline{active}{1d}{q4:=subs(m = me, l1, m*sqrt(-alpha\symbol{94}2+1));
}{}
\end{mapleinput}
\mapleresult
\begin{maplelatex}
 \mapleinline{inert}{2d}{q4 := .51098533412599135054}{}{\[\displaystyle {\it q4}\, := \, 0.51098533412599135054\]}
\end{maplelatex}
\end{maplegroup}
\begin{maplegroup}
\begin{Maple Normal}{The differences in these respective energies give the energy shifts due to the finite size effect.}\end{Maple Normal}

\end{maplegroup}
\begin{maplegroup}
\begin{mapleinput}
\mapleinline{active}{1d}{q5:=Emu-q3;
}{}
\end{mapleinput}
\mapleresult
\begin{maplelatex}
 \mapleinline{inert}{2d}{q5 := 4.412311497*10^(-8)}{}{\[\displaystyle {\it q5}\, := \, 0.000000044123114970000000000\]}
\end{maplelatex}
\end{maplegroup}
\begin{maplegroup}
\begin{mapleinput}
\mapleinline{active}{1d}{q6:=Ee-q4;
}{}
\end{mapleinput}
\mapleresult
\begin{maplelatex}
 \mapleinline{inert}{2d}{q6 := 5.02106*10^(-15)}{}{\[\displaystyle {\it q6}\, := \,{ 5.0210600000000000000\times 10^{-15}}\]}
\end{maplelatex}
\end{maplegroup}
\begin{maplegroup}
\begin{Maple Normal}{The ratio of these shifts can be compared to the Friar result above. The difference is in the 6th digit, which corresponds to about the accuracy we have gone.}\end{Maple Normal}

\end{maplegroup}
\begin{maplegroup}
\begin{mapleinput}
\mapleinline{active}{1d}{q5/q6*(me/mmu)\symbol{94}3;
}{}
\end{mapleinput}
\mapleresult
\begin{maplelatex}
 \mapleinline{inert}{2d}{.99407472561492135455}{}{\[\displaystyle  0.99407472561492135455\]}
\end{maplelatex}
\end{maplegroup}
\begin{maplegroup}
\begin{Maple Normal}{As a test of our numerical integration we can obtain \mapleinline{inert}{2d}{f(r)}{$\displaystyle f \left( r \right) $}
for the point charge case. The point charge equation is \mapleinline{inert}{2d}{e4}{$\displaystyle {\it e4}$}
, but Maple finds this too singular to integrate from zero. Therefore we obtain a series expansion around zero and then use that to set initial conditions slightly away from zero. We use the series solution that behaves like \mapleinline{inert}{2d}{r^sqrt(-alpha^2+1)}{$\displaystyle {r}^{ \sqrt{-{\alpha}^{2}+1}}$}
near the origin rather than the one that behaves like \mapleinline{inert}{2d}{r^(-sqrt(-alpha^2+1))}{$\displaystyle {r}^{- \sqrt{-{\alpha}^{2}+1}}$}
.}\end{Maple Normal}

\end{maplegroup}
\begin{maplegroup}
\begin{mapleinput}
\mapleinline{active}{1d}{Order := 4:
}{}
\end{mapleinput}
\begin{mapleinput}
\mapleinline{active}{1d}{dsolve(e4, \{f(r)\}, series):
subs(_C1 = 0, _C2 = 1, 
e6 := convert(rhs(
}{}
\end{mapleinput}
\end{maplegroup}
\begin{maplegroup}
\begin{Maple Normal}{There is no shooting needed here since we know \mapleinline{inert}{2d}{E}{$\displaystyle E$}
. Using \mapleinline{inert}{2d}{E = q4}{$\displaystyle E={\it q4}$}
for the electron:}\end{Maple Normal}

\end{maplegroup}
\begin{maplegroup}
\begin{mapleinput}
\mapleinline{active}{1d}{l2 := E = q4: ip := 10\symbol{94}(-5):
subs(m = me, l1, l2, e6):
ic := \{f(ip) = subs(r = ip, 
}{}
\end{mapleinput}
\begin{mapleinput}
\mapleinline{active}{1d}{eq := \{subs(m = me, l1, l2, e4)\}:
s1 := dsolve(eq union ic, \{f(r)\}, type = numeric, st1):
odeplot(s1, [[r, f(r)], 
[r, subs(m = me, l1, r\symbol{94}sqrt(-alpha\symbol{94}2+1)*exp(-m*alpha*r))]],
ip .. 9000);
}{}
\end{mapleinput}
\mapleresult
\mapleplot{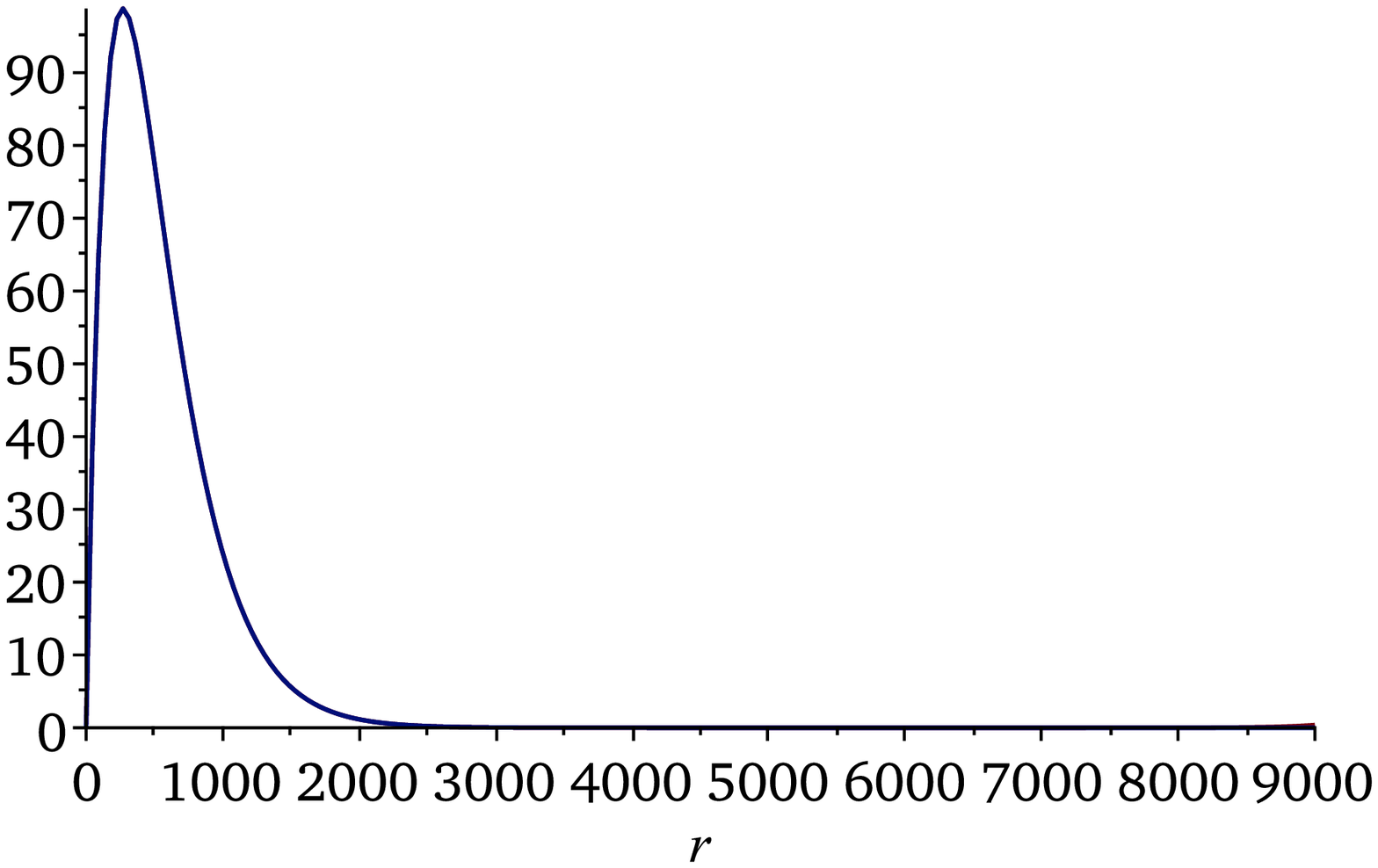}
\end{maplegroup}
\begin{maplegroup}
\begin{Maple Normal}{We have plotted the resulting numerical \mapleinline{inert}{2d}{f(r)}{$\displaystyle f \left( r \right) $}
along with the exact result and the agreement is excellent. The same works for the muon.}\end{Maple Normal}

\end{maplegroup}
\begin{maplegroup}
\begin{mapleinput}
\mapleinline{active}{1d}{l2 := E = q3: ip := 10\symbol{94}(-5):
subs(m = mmu, l1, l2, e6):
ic := \{f(ip) = subs(r = ip, 
}{}
\end{mapleinput}
\begin{mapleinput}
\mapleinline{active}{1d}{eq := \{subs(m = mmu, l1, l2, e4)\}:
s1 := dsolve(eq union ic, \{f(r)\}, type = numeric, st1):
odeplot(s1, [[r, f(r)],
[r, subs(m = mmu, l1, r\symbol{94}sqrt(-alpha\symbol{94}2+1)*exp(-m*alpha*r))]],
ip .. 38);
}{}
\end{mapleinput}
\mapleresult
\mapleplot{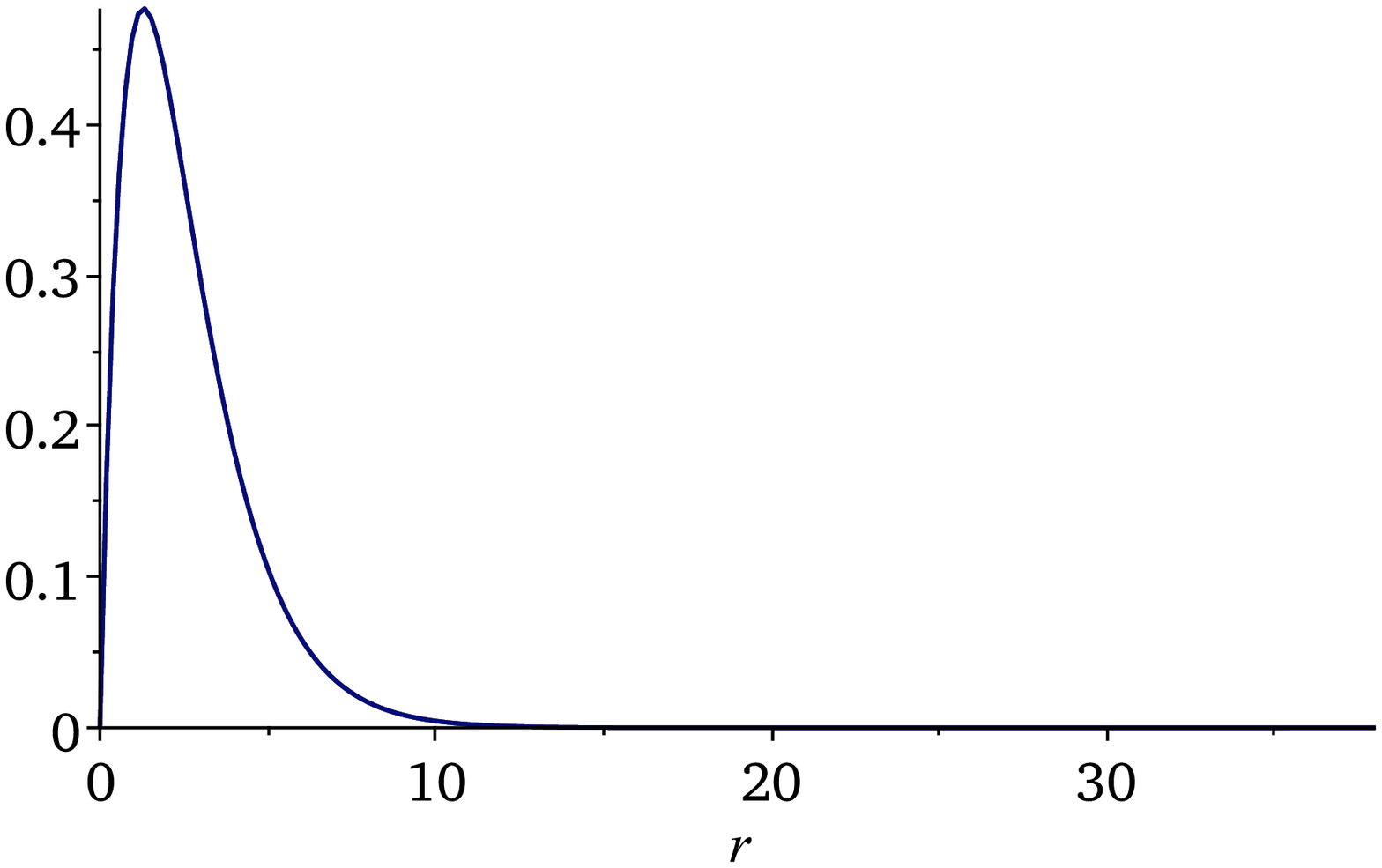}
\end{maplegroup}
\begin{maplegroup}
\begin{Maple Normal}{By using the point charge case as a check we could increase the accuracy of the calculations and push the result for the residual mass dependence beyond 6 digits. But already we see that Friar's perturbative calculations are very accurate.}\end{Maple Normal}

\end{maplegroup}

\section{Nuclear-size effect and the Coulomb Klein-Gordon equation }

Due to some current misconceptions, we present an extended aside on the application of first-order perturbation theory to the nuclear-size effect within the Coulomb Klein-Gordon equation.

The stationary Klein-Gordon Coulomb equation can be written as
\begin{equation}
[\nabla^2 +U(r) +k^2]\psi =0
\end{equation}
where $k^2 = \omega^2-m^2$ and
\begin{equation}
U(r) = 2\omega A_0 - A_0^2 = -\frac{2\omega\alpha}{r}-\frac{\alpha^2}{r^2}
\end{equation}
The energy eigenvalues for the Klein-Gordon Coulomb equation are
\begin{equation}
\omega_{nl} = \frac{m}{\sqrt{1+\displaystyle\frac{\alpha^2}{(n-l+1/2+\sqrt{(l+1/2)^2-\alpha^2)^2}}}}
\end{equation}
A solution to the radial $l=0$ equation is $W(\lambda,\mu,\beta r)/r$ where $W(\lambda,\mu,\beta r)$ is the Whittaker function and 
\begin{eqnarray}
\lambda &=& \alpha \omega/\sqrt{m^2-\omega^2}\nonumber\\
\mu &=& \sqrt{1/4-\alpha^2}\nonumber\\
\beta &=& 2\sqrt{m^2-\omega^2}
\end{eqnarray}

We can rewrite the Klein Gordon equation as a Schr\"odinger-type equation with $V(r) = -\alpha/r$
\begin{equation}
\left[-\frac{\nabla^2}{2m}  +\tilde{U}(r) \right]\psi = \epsilon\psi
\end{equation}
where $\epsilon = \omega-m$, $\omega+m \approx 2m$ and
\begin{equation}
\tilde{U}(r)= -\frac{\alpha}{r}-\frac{\alpha^2}{2mr^2} =V(r)-\frac{V^2(r)}{2m}
\end{equation}
We will now change the short-range potential to $V_{core}(r)$ and assume it is produced by a spherical charge of radius $a$. Thus
\begin{equation}
V_{core}(r)= \frac{\alpha}{2a}\left[\left(\frac{r}{a}\right)^2-3\right]
\end{equation}
The perturbation is therfore
\begin{equation}
\tilde{U}(r)_{pert} = V_{core}-\frac{V^2_{core}(r)}{2m} - V(r)+\frac{V^2(r)}{2m}
\end{equation}
The first-order perturbative correction is given by
\begin{equation}
\Delta E_1 = N^2\int_0^a \tilde{U}_{pert}(r)W(\lambda,\mu,\beta r)^2 \,dr
\end{equation}
where $N$ is a normalization constant.

This integral can be done exactly but yields an extremely long expression. By expanding out the resulting Gamma functions $\Gamma(s)$, and incomplete Gamma functions $\Gamma(s,x)$, and keeping only the leading terms, an excellent approximation $\Delta E_1^a\approx \Delta E_1$ is obtained:
\begin{eqnarray}
\Delta E_1^a &=&\frac{am^2\alpha^4}{29400}\times\Big[48\alpha   (630+(4807-1260\gamma) \alpha ^2)\nonumber\\
&&+49\, am \left(240+(1201-480\gamma) \alpha^2\right)\nonumber\\
&&+560\, \alpha ^2\log(2a m\alpha) (7 a m (5 a m \alpha -6)-108 \alpha)\Big]
\end{eqnarray}
The dominant two terms in this expression are $\Delta E_1^a \approx 36/35\,am^2\alpha^5 + 2/5\,a^2m^3\alpha^4 $ (c.f. Dirac  $2/5\,a^2m^3\alpha^4$. Note that the first term dominates in the electron case and that the second term dominates in the muon case.)

Note that if one doesn't assume that $\omega+m \approx 2m$, but writes 
\begin{equation}
\tilde{U}(r)_{pert}(r) = \frac{\omega}{m}V_{core}-\frac{V^2_{core}(r)}{2m} - \frac{\omega}{m}V(r)+\frac{V^2(r)}{2m}
\end{equation}
one obtains an {\it additional} higher-order effect of
\begin{equation}
\delta\Delta E_1^a = -\frac{1}{5}\,a^2\alpha^6m^3
\end{equation}
{\it i.e.}
\begin{eqnarray}
\Delta E_1^a &=&\frac{am^2\alpha^4}{29400}\times\Big[48\alpha   (630+(4807-1260\gamma) \alpha ^2)\nonumber\\
&&+49\, am \left(240+(1081-480\gamma) \alpha^2\right)\nonumber\\
&&+560\, \alpha ^2\log(2a m\alpha) (7 a m (5 a m \alpha -6)-108 \alpha)\Big]
\end{eqnarray}

\appendix\section{Nuclear-size corrections for a general charge distribution}
Friar \cite{Friar:1978wv} finds
\begin{equation}
\Delta E_n =\frac{2\pi}{3}|\phi_n(0)|^2Z\alpha\left(\langle r^2\rangle - \frac{Z\alpha\mu}{2}\langle r^3\rangle_{(2)} +(Z\alpha)^2F_{REL}+(Z\alpha\mu)^2F_{NR}\right)
\end{equation}
 where
 \begin{equation}
\langle r^p\rangle_{(2)}  = \int d^3s\,d^3r\,\rho(r)\rho(s)|{\bf r-s}|^p
\end{equation}
\begin{equation}
F_{REL} =-\langle r^2\rangle(\langle \log(\beta r)\rangle+\psi(n)+2\gamma-2 ) - \frac{\langle r^3\rangle \langle 1/r\rangle}{3}+ I_2^{REL}+ I_3^{REL}
\end{equation}
\begin{eqnarray}
F_{NR} &=& \frac{\langle r^4\rangle}{10} + \frac{2}{3}\langle r^2\rangle \langle r^2\log(\beta r)\rangle +\frac{2}{3}\langle r^2\rangle^2(\psi(1) + 2\gamma - \frac{7}{3})\nonumber\\
& & + \langle r^3\rangle\langle r\rangle + \langle r^5\rangle\langle1/r\rangle + I_2^{NR} + I_3^{NR}\
\end{eqnarray}
\begin{equation}
I_2^{} = \int d^3s\,\rho(s)\int d^3t\,\rho(t) J^{(2)}_{}(s,t)\Theta(s-t)
\end{equation}
\begin{equation}
I_3^{} = \int d^3u\,\rho(u)\int d^3t\,\rho(t)\int d^3s\,\rho(s)( J^{(3)}_{}(s,t,u)\Theta(u-t)\Theta(t-s) + {\rm sym.})
\end{equation}
\begin{equation}
J^{(2)}_{REL}(s,t) = -(t^2 +s^2)\ln(s/t) - \frac{t^3}{3s} + \frac{s^3}{3t} + \frac{s^2-t^2}{3}
\end{equation}
\begin{eqnarray}
J^{(3)}_{REL}(s,t,u) &=& -\frac{s^2}{3}\ln(s/t)-\frac{s^4}{45tu}+ \frac{s^3}{9}\left(\frac{1}{u} + \frac{1}{t}\right) + \frac{s^2t^2}{36u^2} - \frac{2s^2t}{9u} +\frac{s^2}{9}
\end{eqnarray}
\begin{equation}
J^{(2)}_{NR}(s,t) = \frac{t^5}{9s}-\frac{s^5}{9t}+t^3s -s^3t + \frac{(s^4-t^4)}{2} + \frac{2s^2t^2}{3}\log(s/t)
\end{equation}
\begin{eqnarray}
J^{(3)}_{NR}(s,t,u) &=& \frac{2s^2tu}{3} + \frac{s^4u}{15t} - \frac{s^3u}{3} +\frac{2s^2t^3}{27u} +\frac{s^4t}{15u} +\frac{8s^6}{945tu} -\frac{s^5}{27u}\nonumber\\
& & -\frac{2s^2t^2\log(t/u)}{9} +\frac{2s^2t^2}{27} -\frac{s^3t}{3} -\frac{s^5}{27t} +\frac{s^4}{6}\
\end{eqnarray}
$|\phi_n(0)|^2\equiv (Z\alpha\mu)^3/\pi n^3 $,  $\beta = 2Z\alpha\mu/n$, $\psi(n)$ is the digamma function and $\gamma$ is Euler's constant.

For completeness we give the correction due to recoil when keeping a finite nuclear mass.
\begin{equation}
\Delta E_R = -\frac{(Z\alpha)^4\mu^2}{8m_N}-\frac{(Z\alpha)^5\mu^3}{8m_N}\langle r\rangle_{(2)} +\Delta E_R^{NB}.
\end{equation}
$\langle r\rangle_{(2)} = \frac{36}{35}R$ for the uniform sphere. $\Delta E_R^{NB}$, the ``non-Breit" finite size correction of order $(Z\alpha)^5$, is expected to be small.


\begin{thebibliography}{99}

\bibitem{BandD} 
  J.~D.~Bjorken and S.~D.~Drell,
  ``Relativistic quantum mechanics,'' McGraw-Hill, New York, 1964, ISBN-0070054932.
  
\bibitem{Friar:1978wv} 
  J.~L.~Friar,
  Annals Phys.\  {\bf 122}, 151 (1979).

\end{thebibliography}
\end{document}